\documentclass[3p,times,procedia]{elsarticle}
\usepackage{nupha_ecrc}
\usepackage{amsmath}
\usepackage{subcaption}
\usepackage{adjustbox}

\volume{00}

\firstpage{1}

\journalname{Nuclear Physics A}

\runauth{}

\jid{nupha}

\jnltitlelogo{Nuclear Physics A}

\usepackage{amssymb}

\usepackage[figuresright]{rotating}

\usepackage{lineno}

\begin{document}

\begin{frontmatter}




\title{Confronting fluctuations of conserved charges in central nuclear collisions at the LHC with predictions from \\ Lattice QCD}


\address[EMMI]{Extreme Matter Institute EMMI, GSI,Planckstr. 1, D-64291 Darmstadt, Germany}
\address[TUD]{Technische Universit\"at Darmstadt, Darmstadt, Germany}
\address[FIAS]{Frankfurt Institute for Advanced Studies, J.W. Goethe Universit\"at, Frankfurt, Germany}
\address[CERN]{European Organization for Nuclear Research (CERN), Geneva, Switzerland}
\address[WROCLAW]{Institute of Theoretical Physics, University of Wroclaw,
PL-50204 Wroc\l aw, Poland}
\address[HD]{Physikalisches Institut, Universit\"at Heidelberg, Heidelberg, Germany}

\author[EMMI,TUD,FIAS,HD]{P. Braun-Munzinger}
\author[CERN]{A. Kalweit}
\author[WROCLAW,EMMI]{K. Redlich}
\author[HD]{J. Stachel}


\address{}

\begin{abstract}

We construct net baryon number and strangeness susceptibilities as well as correlations between electric charge, strangeness and baryon number from experimental data on the particle production yields at midrapidity of the ALICE Collaboration at CERN. The data were taken in central Pb-Pb collisions at $\sqrt{s_{\rm NN}}$~=~2.76~TeV and cover one unit of rapidity. We show that the resulting fluctuations and correlations are consistent with Lattice QCD results at the chiral crossover pseudocritical temperature $T_{c} \simeq$ 155 MeV. This agreement lends strong support to the assumption that the fireball created in these collisions is of thermal origin and exhibits characteristic properties expected in QCD at the transition from the quark gluon plasma to the hadronic phase. Since Lattice QCD calculations are performed at a baryochemical potential of $\mu_{B}$ = 0, the comparisons with LHC data are the most direct due to the vanishing baryon transport to midrapidity at these high energies.

\end{abstract}


\end{frontmatter}


\section{Introduction and motivation}
\label{sec.:Introduction}

Particle production in ultra-relativistic heavy-ion collisions is a complex process and its theoretical characterisation combines fundamental aspects of Quantum Chromodynamics (QCD) with many-body theories which describe the produced matter in thermal equilibrium. Ultimately, the goal is an as complete description as possible based on ab-initio QCD calculations without any further modelling or free parameters. Ab-initio calculations are only feasible in certain regimes of the QCD phase diagram. If the chemical potentials $\mu$ of the conserved charges in QCD (baryon number $B$, strangeness $S$, and electric charge $Q$) and the temperature $T$ are small with respect to the QCD scale parameter $\Lambda_{QCD}$, i.e. $\mu , T  \ll \Lambda_{QCD}$, then approaches based on chiral perturbation theory are applicable. In the opposite case ($\mu , T  \gg \Lambda_{QCD}$), the running coupling allows perturbative QCD calculations. For the production of the bulk of light flavour hadrons, i.e. those hadrons which contain only $u$, $d$, and $s$ valence quarks, both approaches are not suited. About 98\% of all particles at LHC energies are produced with transverse momenta $p_{\rm T} < 2$~GeV/$c$, thus in a non-perturbative regime. Consequently, the only available ab-initio calculations are based on lattice gauge theory \cite{Allton:2005gk,Borsanyi:2011sw,Bazavov:2012jq} which is applicable for $\mu \ll T$. As a matter of fact, this condition is best fulfilled at LHC energies where the temperature is high and the baryon number transport to midrapidty is negligible, thus allowing for the most direct comparison between theory and experiment.

From the experimental side, the complete set of light flavour hadron particle yields has been measured by the ALICE collaboration in the recent years \cite{Abelev:2013vea,Abelev:2013xaa,ABELEV:2013zaa,Abelev:2014uua}. It was found that the thermal model describes the data on all particle yields within $3\sigma$ \cite{Floris:2014pta,BraunMunzinger:2003zd}. Nevertheless, one of the remaining open issues is the question if the quality of the fit can be further improved by including feed-down from missing resonances. As unknown resonances are naturally included in lattice QCD (LQCD) calculations, it is of major importance to confirm the consistency of LQCD calculations with the measured particle yields. The approach presented here, which was published in detail in \cite{Braun-Munzinger:2014lba}, provides this connection and allows a direct comparison of LQCD results with experimental data on particle yields measured at the LHC.

\section{Construction of conserved charge susceptibilities from particle yields}
\label{sec.:Construction}

In the grand-canonical ensemble, the conservation laws are ensured by the chemical potentials $\vec{\mu} = (\mu_B, \mu_S, \mu_Q)$. Fluctuations ${\hat{\chi}_N}$ and correlations ${\hat{\chi}_{NM}}$ of conserved quantities are quantified by the susceptibilities 
		
		\begin{equation}
			{\hat{\chi}_N} = {\chi_N \over T^{2}} = {\partial^{2}\hat{P} \over \partial\hat{\mu}_N^2} \;\;\;\;\;
			{\hat{\chi}_{NM}}  = {\chi_{NM} \over T^{2}} = {\partial^{2}\hat{P} \over \partial\hat{\mu}_N\partial\hat{\mu}_M} \;\;,
		\end{equation}
		
\noindent where $\hat{P} = P/T^{4}$ corresponds to the reduced pressure and $\hat{\mu}=\mu/T$ to the reduced chemical potential of the conserved quantity $N,M = (B,S,Q)$. The susceptibility of a conserved charge can be also related to its variance ${\hat{\chi}_N} = {1 \over VT^3} \Bigl(\langle N^2 \rangle - \langle N \rangle^2 \Bigr)$,
where $N = N_q - N_{-q}$ corresponds to the difference between the number of particles with charge $q$ and those with charge $-q$ and $V$ to the volume. With the probability distribution $P(N)$, the $n$-th moment $\langle N \rangle$ is calculated as $\langle N^n\rangle = \sum N^n P(N)$. In the study presented here, the baseline assumption is that charge $N_q$ and anti-charge $N_{-q}$ are uncorrelated and follow a Poisson distribution. Then it follows that $P(N)$ is given by the so-called Skellam distribution:

\begin{equation}
	P(N) = \Bigl({\langle N_q \rangle  \over \langle N_{-q} \rangle} \Bigr)^{N/2} \, I_{N}\bigl(2\sqrt{\langle N_q \rangle \langle N_{-q} \rangle}\bigr) \exp[-(\langle N_{q} \rangle +\langle N_{-q} \rangle)] \;\; .
\end{equation}		

\noindent In the following, we make use of the feature that the variance of the Skellam distribution can be directly calculated from the mean number of particles and anti-particles as $\langle N^2 \rangle - \langle N \rangle^2 =  \langle N_q \rangle +  \langle N_{-q} \rangle$.
And thus the susceptibility is obtained as

\begin{equation}
	{\hat{\chi}_N}  = {1 \over VT^3} \Bigl( \langle N_q \rangle +  \langle N_{-q} \rangle \Bigr)	\; .
\end{equation}

\noindent However, the previous equation is only valid if there are only particles of the same charge, as for the baryon number, where the charge is $B=\pm1$. For strangeness and electric charge, there are hadrons with charge two and three. In this case, the Skellam probability distribution has to be generalised and results in the following expressions for the susceptibilities \cite{BraunMunzinger:2011ta}:


\begin{equation}
 \hat\chi_N = {{\chi_N}\over T^2}=\frac{1}{VT^3}\sum_{n=1}^{|q|} n^2(\langle N_{n}\rangle+\langle N_{-n}\rangle) \;\;\;\;\;\;
 \hat\chi_{NM} = {{\chi_{NM}}\over T^2}=\frac{1}{VT^3}\sum_{n=-q_N}^{q_N} \sum_{m=-q_M}^{q_M}n  m\langle N_{n,m}\rangle 
\end{equation}

\noindent where  $\langle N_{n,m}\rangle$, is the mean number of particles and resonances  carrying charges $N=n$ and $M=m$.

\section{Results}
\label{sec.:Results}

\begin{figure}
\centering
 \includegraphics[width=0.43\textwidth]{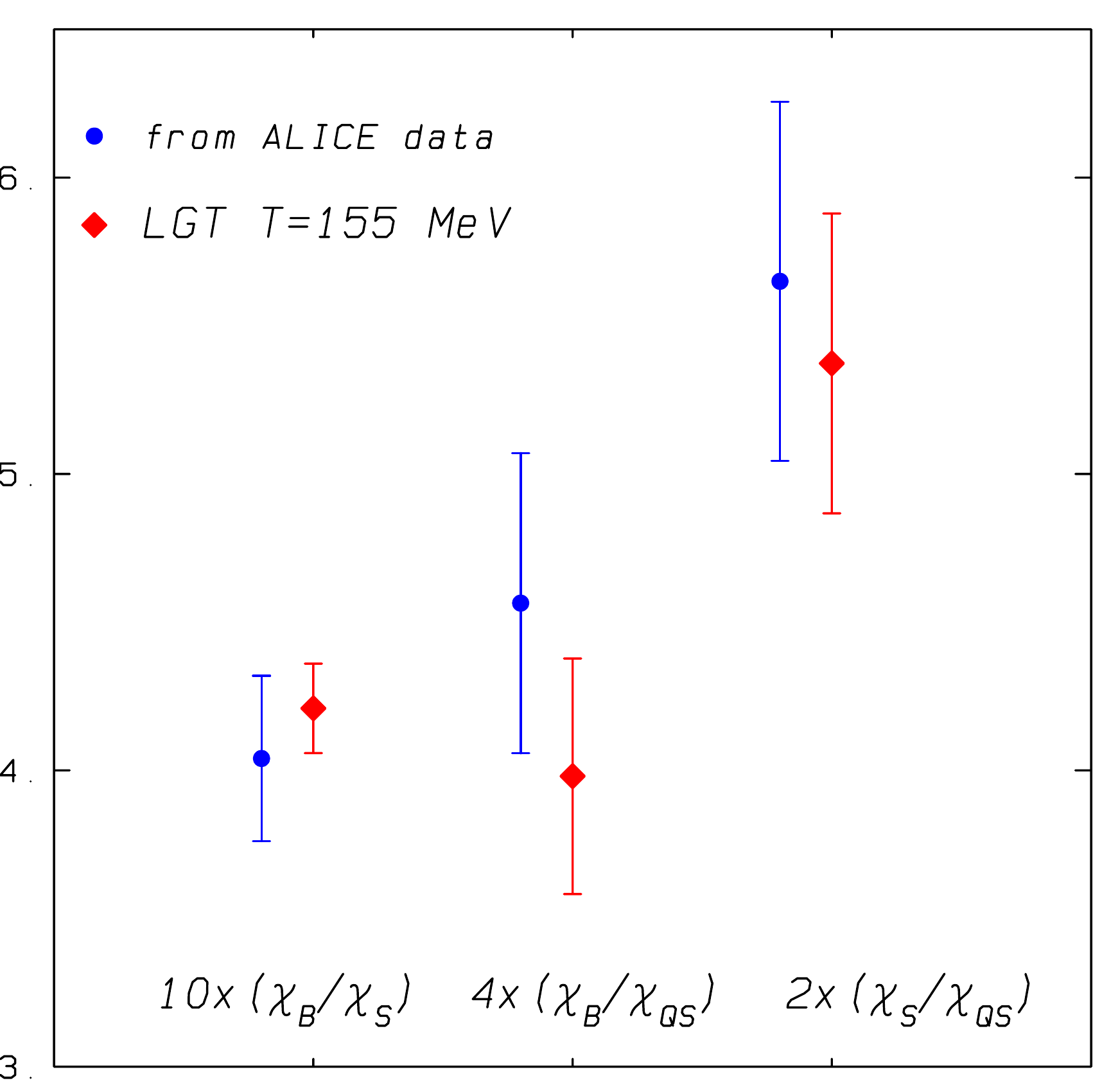}
\hspace{1cm}
 \includegraphics[width=0.43\textwidth]{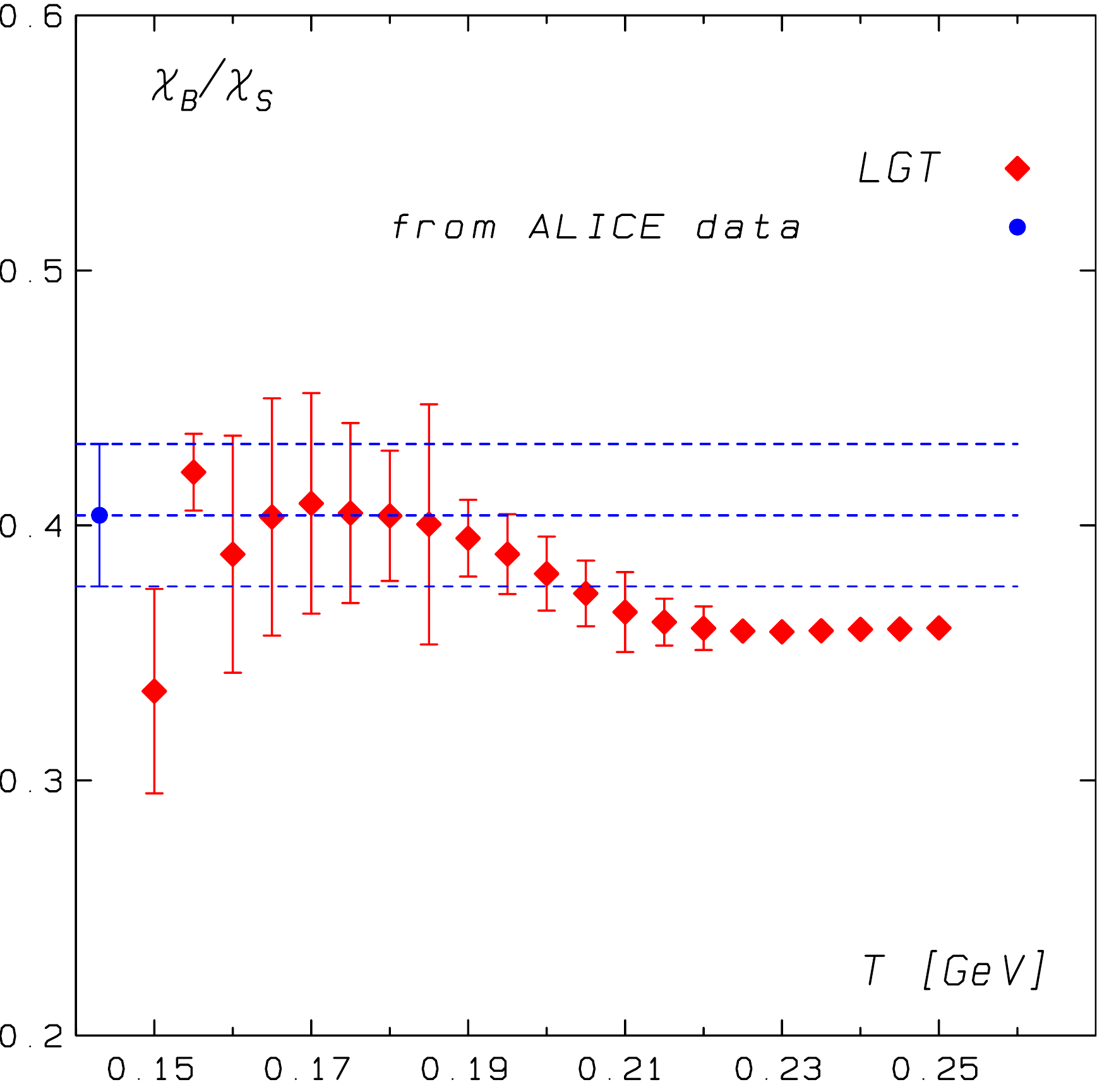}
   \caption{Comparison of different susceptibility ratios obtained from Eqs. (\ref{eq.:chiB}), (\ref{eq.:chiS}), and (\ref{eq.:chiQS}) by using data measured by the ALICE collaboration with LQCD results  from Refs. \cite{Bazavov:2012jq,Bazavov:2014xya} at $T=155$~MeV (left) and as a function of temperature for $\chi_B/\chi_S$ (right). The LQCD value at $T=155$ MeV is from Ref. \cite{Bazavov:2014xya}. }
   \label{fig1}
 \end{figure}

After some further assumptions (see \cite{Braun-Munzinger:2014lba} for details) and inserting the values for particle yields measured by the ALICE collaboration \cite{Abelev:2013vea,Abelev:2013xaa,ABELEV:2013zaa,Abelev:2014uua}, the following results are obtained:

\begin{align} \label{eq.:chiB}
{{\chi_B}\over T^2} &= {1\over{VT^3}}[  4\langle p\rangle+ 2\langle(\Lambda+\Sigma^0)\rangle + 4\langle\Sigma^{+}\rangle   +4\langle\Xi\rangle +2\langle\Omega\rangle] 
=  {1\over{VT^3}} (203.7\pm 11.44)  \\\nonumber \\\nonumber
 {{\chi_S}\over T^2} &\simeq {1\over{VT^3}}[ 2 \langle K^{+}\rangle+2\langle K^0\rangle
+2\langle(\Lambda+\Sigma^{0})\rangle
  +4\langle\Sigma^{+}\rangle+16\langle\Xi\rangle+
18\langle\Omega\rangle
- 2(\Gamma_{\phi\to K^+}+\Gamma_{\phi\to K^0})\langle\phi\rangle ] \\
&= {1\over{VT^3}}(504.35\pm 24.14) \label{eq.:chiS}  \\\nonumber \\
 \label{eq.:chiQS} {{\chi_{QS}}\over T^2} &\simeq {1\over{VT^3}}[ 2 \langle K^{+}\rangle
+4\langle\Xi^{-}\rangle+
6\langle\Omega^{-}\rangle
- 2\Gamma_{\phi\to K^+}\langle\phi\rangle - 2\Gamma_{K_0^*\to K^+}\langle K^*\rangle] 
= {1\over{VT^3}}( 178.5\pm 17.14) 
\protect
\end{align}

\noindent As can be seen, in the case of strange particles, contributions of particles with $S=0$ decaying into particles with $S\neq0$ need to be subtracted. Furthermore, the production yield of protons and neutrons is assumed to be equal which is also experimentally justified based on the fact that no surprises were found in the measurement of the production yields of (anti)-deuterons or heavier (anti-)nuclei \cite{nuclei}.

\begin{figure}
\centering
  \includegraphics[width=0.43\textwidth]{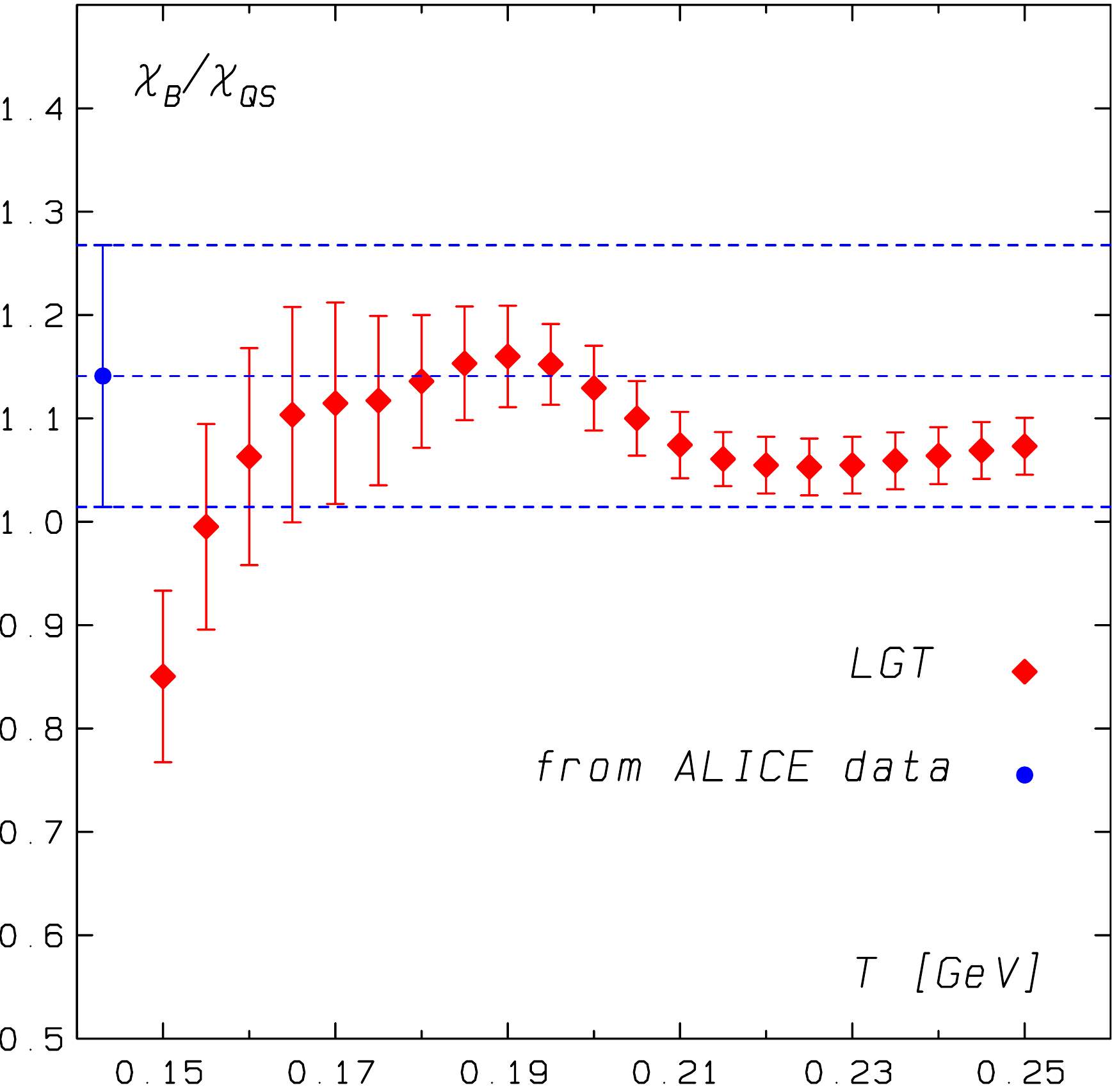}
  \hspace{1cm}
  \includegraphics[width=0.43\textwidth]{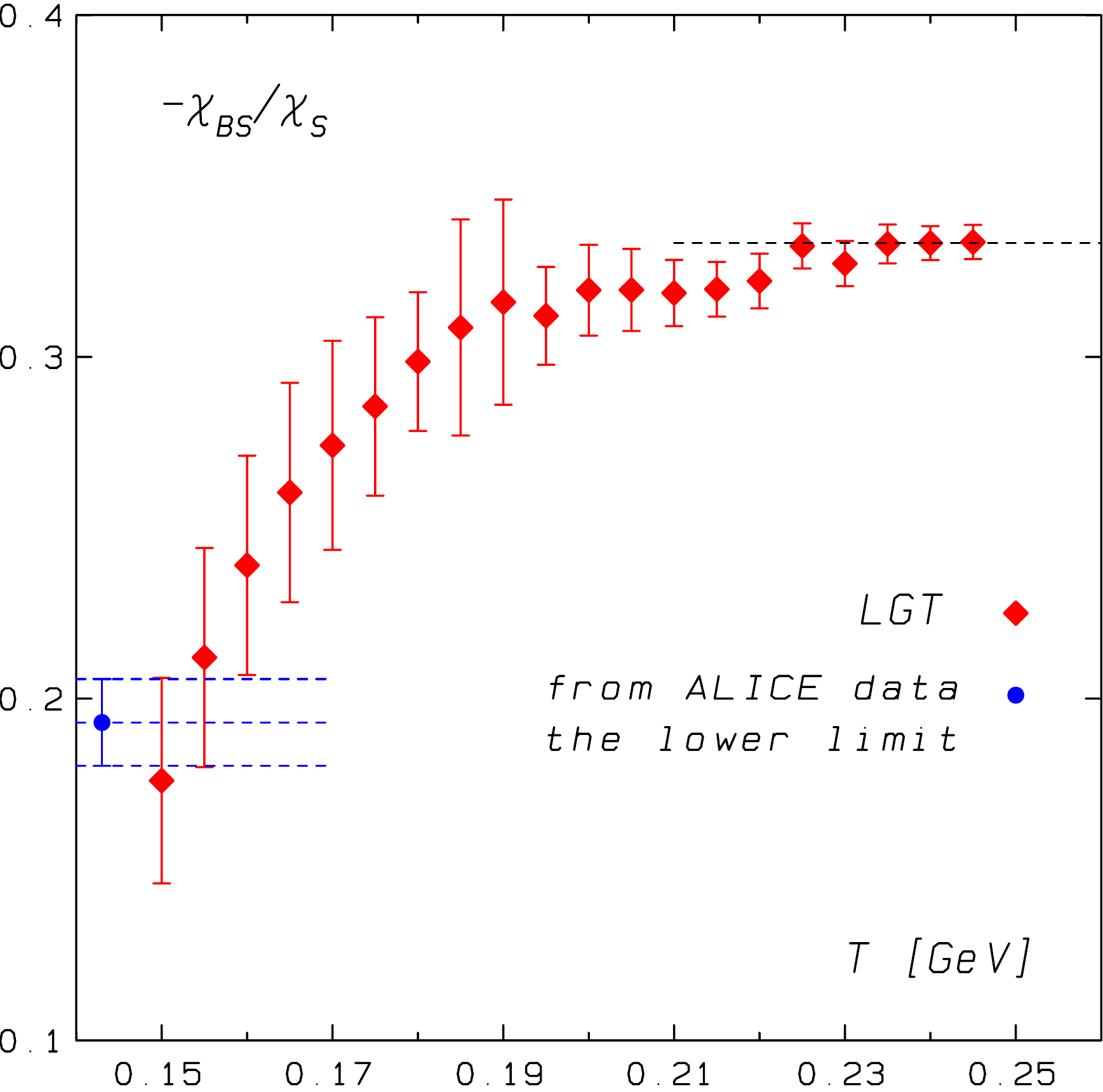}
  \caption{Ratio $\chi_B/\chi_{QS}$ (left) and $-\chi_{BS}/\chi_{S}$ (right) from LQCD data from
   Refs. \cite{Bazavov:2014xya,Bazavov:2012jq}, and obtained from ALICE data based on Eqs.~(\ref{eq.:chiB}), (\ref{eq.:chiS}), and (\ref{eq.:chiQS}) shown as a blue band.   The horizontal line at high-$T$ is an ideal gas value in  a QGP.}\label{fig2}
 \end{figure}

For a comparison with the LQCD results, ratios of the susceptibilities are formed in which the dependence on the volume cancels. If there is a phase change from QGP to the hadronic phase, particle yields and fluctuations of conserved charges are expected to be established at the chiral, pseudocritical temperature $T_c$. A value of $T_c = 155(1)(8)$ MeV was recently obtained in LQCD and, as shown in the left panel of Fig.~\ref{fig1}, a very good agreement is found between susceptibilities derived from ALICE particle yield data and LQCD results at this temperature. Unfortunately, the temperature at which this agreement holds cannot be determined uniquely. As shown in the right panel of Fig.~\ref{fig1} and the left panel of Fig.~\ref{fig2}, the ratios exhibit only a weak dependence on temperature. Nevertheless, the temperature range $T < 150$~MeV is rejected as a saturation regime for $\chi_B$, $\chi_{S}$, and $\chi_{QS}$. As a matter of fact, the strongest temperature dependence is visible in the ratio $-\chi_{BS}/\chi_{S}$. Unfortunately, only a lower limit can be set for this ratio, because contributions of strange baryonic resonances decaying into a non-strange baryon and a strange meson (e.g. the decay of  $\Sigma^*\rightarrow N\bar{K}$ ) are not determined experimentally. Similarly to the other ratios, also the lower limit on $-\chi_{BS}/\chi_{S}$ excludes temperatures below 150~MeV. At the same time, based on different combinations of charge fluctuations and correlations, it was shown, that at $T>163$ MeV, the LQCD
thermodynamics can not be anymore described by hadronic degrees of freedom \cite{Karsch:2013naa}. This argument reduces a conceivable window for the saturation of the net baryon number and strangeness fluctuations to $0.15<T<0.163$ GeV.

\section{Summary and conclusions}
\label{sec.:Summary}

We found direct agreement between susceptibilities derived from experimentally measured particle yields and LQCD calculations in the temperature regime of the chiral crossover $T_{c} \approx 155$~MeV. This observation lends strong support to the notion that the fireball created in central nucleus-nucleus collisions at the LHC is of thermal origin and exhibits characteristic properties expected in QCD at the transition from a quark-gluon plasma to a hadronic phase. The analysis presented here provides the first direct link between LHC heavy ion data measured in the full momentum range and predictions from LQCD. Actual measurements of event-by-event fluctuations of net charges at the LHC are eagerly awaited in order to test if the assumption that the variance is described by the Skellam distribution is valid. Furthermore, the study presented here also provides a baseline for the investigation of deviations from the Skellam distribution for higher order moments ($n  \geq 6$) caused by remnants of the critical chiral dynamics \cite{Friman:2011pf} at $\mu \approx 0$.




\vspace{-0.3cm}

\bibliographystyle{elsarticle-num}
\bibliography{biblAK}







\end{document}